\documentclass[twocolumn,amsmath,amssymb]{snp}
\usepackage{graphicx}
\usepackage{dcolumn}
\usepackage{bm}
\begin{document}

\title{{Spectroscopy of di-meson bound states in charm and beauty sector}}

\author{\large Smruti Patel}
\email{fizix.smriti@gmail.com}
\author{\large Manan Shah}
\author{\large Arpit Parmar}
\author{\large P.C.Vinodkumar}\email{p.c. vinodkumar@gmail.com}
\affiliation{Deptartment of Physics, Sardar Patel University,Vallabh Vidyanagar-388120, Gujarat, INDIA}

\maketitle

\section*{Introduction}
Very recently there exists increasing attention towards the study of four quark states as di-hadronic molecular states followed by the recent discovery of $Z_c$(3900) state by two seperate experimental groups BES III \cite{Ablikim} and BELLE Collabaration \cite{Liu}. The interpretation of the new state has triggered a considerable amount of theoretical work, especially due to the controversies related their internal structure. Moreover, very recently BELLE Collaboration has made the tantalizing observation of two new charged bottom resonances, namely Z$_b$(10610) and  Z$_b$(10650). Since all the standard bottomonia are neutrally charged, these two resonances have a flavour only compatible with $b\bar{b}u\bar{d}$ tetraquarks \cite{Adachi,Bondar}.\\
Motivated by striking observation of tetraquark states, here we wish to predict the interpretation of these states as di-mesonic molecules composed of a pair of heavy mesons such as $D\bar{D}$, $D\bar{D}^*$, $D^*\bar{D}^*$, $D^+\bar{D}^*$ in the charm sector and $B\bar{B}$, $B^*\bar{B}^*$, $B\bar{B}^*$ in the bottom sector.

\section*{Phenomenology}
Different attempts have been made for the interpretation of exotic hadronic states of four quark system such as di-mesonic states, hadroquarkonium states and tetraquark states. Investigations into the existence of multiquark states have begun in the early days of QCD \cite{Jaffe,Strottman}. However, little success has been achieved in understanding tetraquark states due to the non-perturbative nature of QCD at the hadronic scale. The hadron molecular considerations does simplify this difficulty by replacing interquark color interacion with a residual strong interactions between two color singlet hadrons. Thus, for the present study of di-mesonic molecules, we employ Woods Saxon plus coulomb type of potential between two color singlet hadrons of the form
\begin{equation}\label{eq:1}
V(r)=\frac{V_0}{1+\exp^{(\frac{r-R}{a})}}-\frac{B}{r}
\end{equation}
\begin{table*}
\begin{center}
\tabcolsep 0.002pt
 \small
\caption{Mass spectra of di-mesonic systems (in MeV).}\label{tab1}
\begin{tabular}{|c|c|c|c|c|c|c|c|}
\hline
Molecule	&	$J^{pc}$	 &	BE	&	$E_{(j_1, j_2; J)}$	&	$M_{SA}$	&	$M_J$	& EXP.	 & Others	\\
\hline
&		&		&		&		&		&		& 3738\cite{rai}\\
$D\bar{D}$	&	$0^{++}$	&	28.87	&	0.0	&	3758.87	&	3758.87	&	-	&	3760$\pm100$\cite{Jian}	\\		
	&		&		&		&		&		&		&		$3715^{+24}_{-27}$\cite{Hidalgo}\\

\hline
&		&		&		&		&		&		& 3878[9] \\

$D\bar{D}^*$	&	$1^{++}$	&	28.79	&0.0		&	3900.8	&	3900.8	&	X(3871.68$\pm0.17$)\cite{Beringer}	&		3880$\pm110$\cite{Jian} \\

\hline
&		&		&		&		&		&		&\\
{$D^+\bar{D}^*$}	&	$1^{+-}$	&	28.78	&0.0		&	3904.79	&	3904.79	&	$Z_c^{+}(3898\pm5)$	&	-	\\		
\hline
&		&		&		&		&		&		&\\
$D^*\bar{D}^*$	&	$2^{++}$	&	28.69	&	0.358	&	4042.69	&	4043.05	&	-	&	4062[9], $4012^{+4}_{-9}$\cite{Hidalgo}	\\		
	&	$1^{+-}$	&		&	-0.358	&		&	4042.33	&	-	&	3974\cite{rai}, $3958^{+24}_{-27}$\cite{Hidalgo}	\\		
	&	$0^{++}$&		&	-0.72	&		&	4041.97	&	-	&	3930\cite{rai}	\\		
\hline
&		&		&		&		&		&		&\\
$B\bar{B}$	&	$0^{++}$	&	19.06	&	0.0	&	10577.1	&	10577.1	&	-	&	-	\\		
\hline
&		&		&		&		&		&		&\\
$B\bar{B}^*$	&	$1^{+-}$	&	18.99	&	0.0	&	10623	&	10623	&	$Z_b$(10607$\pm2.0$)\cite{Bondar}	&	-	\\			
\hline
&		&		&		&		&		&		&\\
$B^*\bar{B}^*$	&	$2^{++}$	&	18.92	&1.367		&	10668.9	&	10670.28	&	-	&	-	\\		
	&	$1^{+-}$	&		&-1.367		&	&	10667.53	&	$Z_b$(10652.2$\pm1.5$)\cite{Bondar}	&-		\\		
	&	$0^{++}$	&		&-2.73		&		&	10666.16	&	-	&	-	\\		

\hline

\end{tabular}
\end{center}
\begin{center}
\end{center}
\end{table*}
The potential parameters employed here are as follows: a$=$-0.0387 fm; $V_0$=0.03 GeV; B=0.04; R=0.8875 fm. Binding energy is obtained by numerically solving Schr\"{o}dinger equation using mathematica notebook of Range-Kutta method. The non-relativistic Schrodinger bound-state mass (spin average mass) of the di-mesonic system is obtained as
\begin{equation}\label{eq:3}
M_{SA}=m{_1}+m{_2}+BE
\end{equation}
We introduce j-j coupling term to obtain the hyperfine splitting of the different di-meson states. Accordingly, the di-mesonic molecular mass is obtained as
\begin{equation}\label{3}
M_J=M_{SA}+E_{(j_1, j_2; J)}
\end{equation}
Where m$_1$ and m$_2$ are the masses of the constituent mesons, BE represents the binding energy of the di-mesonic system and $E_{(j_1, j_2; J)}$ represents the spin-dependent term. The hyperfine interaction is computed using the expression similar to the hyperfine interactions for quarkonia but without considering color factor and is taken as
\begin{equation}\label{eq:4}
E_{(j_1, j_2; J)}=\frac{2<j_1.j_2>_J|R(0)|^2}{3m_1m_2}
\end{equation}

\section*{Results and conclusion}

Table \ref{tab1} summarizes the binding energies and  low lying masses of the di-mesonic states. The recent experimental exotic states and other theoretical results are also presented for comparision. In the present work, the mass of $D\bar{D}^*$ is 18 MeV above the experimental value. Overall agreement of the present results gives us a clue to believe that X(3872) state is nothing but the loosely bound $D\bar{D}^*$ meson molecule  and its companion $D\bar{D}$$(0^{++})$ is predicted to be at 3759 MeV. In the same spirit two bottomonium-like twin resonances Z$_b$(10610) and Z$_b$(10650) are found to be $B\bar{B}^*$ and $B^*\bar{B}^*$ molecules respectively. And recent $Z_c^{+}$(3900) state is found to be the $D^+\bar{D}^*$ molecular state. Other positive parity molecular states ($D^*\bar{D}^*)_{J=0,1,2}$ close to $\psi$(4040) are predicted around 4042 MeV. Other di-mesonic molecular states in the charm and beauty sector are also presented in table \ref{tab1}. Many of these states require further experimental support.

\section*{Acknowledgments}
The work is part of a Major research project No. F. 40-457/2011(SR) funded by UGC.

\end{document}